\documentclass{article}

\usepackage{PRIMEarxiv}

\usepackage[utf8]{inputenc} 
\usepackage[T1]{fontenc}    
\usepackage{hyperref}       
\usepackage{url}            
\usepackage{booktabs}       
\usepackage{amsfonts}       
\usepackage{nicefrac}       
\usepackage{microtype}      
\usepackage{lipsum}
\usepackage{graphicx}
\usepackage{subcaption}
\graphicspath{{media/}}     

\title{OFCnetLLM: Large Language Model for Network Monitoring and Alertness
}

\author{
  Hong-Jun Yoon \\
  Oak Ridge National Laboratory \\
  Oak Ridge, TN \\
  \texttt{yoonh@ornl.gov} \\
   \And
  Mariam Kiran \\
  Oak Ridge National Laboratory \\
  Oak Ridge, TN \\
  \texttt{kiranm@ornl.gov} \\
  \AND
  Danial Ebling \\
  Utah Education Network \\
  Salt lake City, Utah \\
  \texttt{danial@uen.org} \\
  \And
  Joe Breen \\
  University of Utah \\
  Salt Lake City, Utah \\
  Joe.Breen@utah.edu \\
}

\begin{document}
\maketitle

\begin{abstract}
The rapid evolution of network infrastructure is bringing new challenges and opportunities for efficient network management, optimization, and security. With very large monitoring databases becoming expensive to explore, the use of AI and Generative AI can help reduce costs of managing these datasets. This paper explores the use of Large Language Models (LLMs) to revolutionize network monitoring management by addressing the limitations of query finding and pattern analysis. We leverage LLMs to enhance anomaly detection, automate root-cause analysis, and automate incident analysis to build a well-monitored network management team using AI. Through a real-world example of developing our own OFCNetLLM, based on the open-source LLM model, we demonstrate practical applications of OFCnetLLM in the OFC conference network. Our model is developed as a multi-agent approach and is still evolving, and we present early results here.

\end{abstract}

\keywords{Deep Learning \and Network Monitoring \and Multi-agent architecture \and Large Language Model Adaptation}

\section{Introduction}
Network service providers are witnessing relentless traffic growth and unique needs for applications, in terms of throughput, loss, security and more. Network engineering and operations are balancing network performance and capacity planning \cite{cisco}, by routinely making active decisions to find efficient network paths to utilize available bandwidth and re-engineering traffic routes for growing traffic. Efficient monitoring and capturing performance data are essential to study and improve these networks. Novel technologies such as programmable network interface cards (NICs) (another term used for SmartNICs) are accelerating packet processing and traffic mirroring to capture petabytes of just network performance data, building big data sets to study \cite{10620201}.

As data increases to vast levels, many networking tasks are exploring the use of artificial intelligence (AI) and machine learning (ML) to help solve complex predictions and find patterns in their data. As networks grow in complexity and scale, the need for a flexible, scalable, and efficient data management solution becomes critical, where traditional data models are becoming cumbersome to monitor, query, and control. For example, multiple databases for Simple Network Management Protocol (SNMP) counters, netflow data, and interface data are examples of the wide variety and uniqueness of each type of data being collected, such that each has unique needs and mechanisms for analysis and data flow management.

The recent emergence of large language models (LLMs) in the field of Generative AI, such as ChatGPT and Llama models, involves computational systems with extensive trainable parameters that demonstrate capabilities analogous to human cognitive functions, encompassing contextual comprehension and logical reasoning to generate coherent responses to complex queries. Through the utilization of extensive datasets and sophisticated architectural frameworks, LLMs have fundamentally transformed the natural language processing (NLP) landscape, facilitating enhanced semantic understanding and natural language generation capabilities. These advancements in LLMs have facilitated the development of AI agent systems. These LLM-based AI agents demonstrate enhanced capabilities in natural language understanding, intent recognition, action execution, and human-computer interaction. This technological progression has resulted in the implementation of AI-assisted systems capable of executing diverse computational tasks, including schedule optimization, recommendation generation, and information retrieval operations. A significant application domain for AI agent systems encompasses the surveillance and analysis of mission-critical systems, where continuous data monitoring and real-time analytical processing are fundamental requirements.

Recent works of using LLMs in networking have argued that these help improve predictions and generalization with powerful pre-trained knowledge of a foundation model that can be ``one model for all tasks'' \cite{10.1145/3651890.3672268}. LLMs are being introduced to compose intent commands to help schedule flows \cite{10574890} or even to configure networks \cite{10.1145/3656296}. Motivated by these works, in this paper, we employ LLMs to design efficient query management systems that can help find patterns and network anomalies and also aid with fault finding. Although LLMs prove to be powerful and reduce the cost of fine-tuning, we find there is a need to efficiently design the architecture with multiple agents that can also impact security and correct reasoning of the data. We discuss early results in this paper.

The paper has been organized in the following manner. Section 2 provides a comprehensive analysis of LLM architectures and examines their algorithmic foundations. Section 3 describes the design and innovation for OFCnetLLM which we present here, with results shown in Section 4. Finally, we present key findings and conclusions in Section 5, synthesizing emerging paradigms within the agentic framework and evaluating their potential implications.

\begin{figure*}
\centering
\includegraphics[height=3in, width=6in]{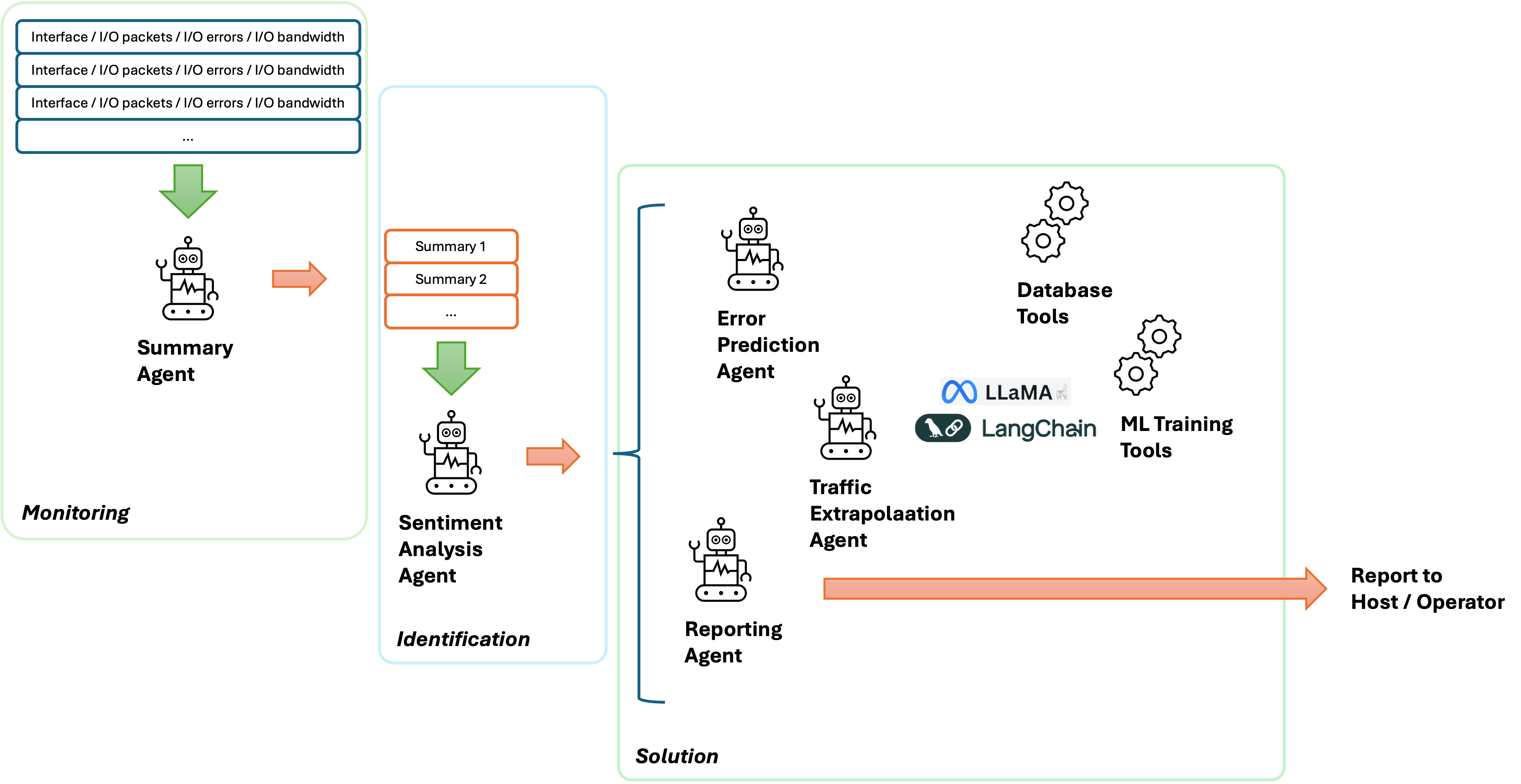}
\caption{Architecture for OFCnetLLM.}\label{arch}
\vskip -6pt
\end{figure*}

\section{LLM Research for Networks}
The integration of AI, ML, and LLMs \cite{Brown_GPT3_2020} with software defined networking (SDN) offers transformative solutions to address network management challenges. These technologies enhance network operations through automated, intelligent decision-making processes that significantly improve both efficiency and security. These include AI/ML algorithms for sophisticated traffic management and resource allocation by analyzing network patterns in real-time, allowing SDN controllers to optimize routing decisions and bandwidth distribution proactively \cite{hecatepaper}, or in the security domain, AI/ML systems detecting anomalies and potential threats by identifying subtle deviations from normal network behavior, enabling rapid response to security incidents. The recent emergence of LLMs has further enhanced network management capabilities by enabling natural language interfaces for network configuration, automated documentation generation, and intelligent troubleshooting assistance. LLMs can interpret complex network logs, suggest optimization strategies, and even generate security policies based on best practices and current threat landscapes.

The emergence of LLMs represents a significant advancement in AI, fundamentally altering the landscape of natural language processing. These models, built upon the Transformer architecture, have evolved from rudimentary neural networks into sophisticated systems capable of comprehending and generating complex linguistic structures. The pioneering models, such as Bidirectional Encoder Representations from Transformers (BERT)\cite{devlin2019bert} and Generative Pre-trained Transformer (GPT)\cite{yenduri2024gpt}, demonstrated exceptional capabilities in natural language understanding and generation through their extensive parametrization and advanced architectural design.

The evolution of LLMs has been characterized by an exponential increase in model complexity and scale. Early transformer-based models comprised millions of parameters, while contemporary models like GPT-4\cite{openai2023gpt4} incorporate hundreds of billions or even trillions of parameters. This substantial increase in scale, combined with advanced training methodologies such as few-shot learning and chain-of-thought prompting, has enabled these models to exhibit remarkable proficiency in tasks ranging from basic text completion to complex reasoning and analysis. LLMs also leverage self-supervised learning at scale. In contrast to traditional supervised learning approaches that necessitate labeled datasets, LLMs are pre-trained on extensive corpora of unlabeled text from diverse sources. This pre-training process facilitates the development of a comprehensive understanding of linguistic patterns, contextual relationships, and domain-specific knowledge. The subsequent fine-tuning process then adapts these pre-trained models for specific tasks or domains, enabling more efficient and effective specialization in various research fields. Additionally, reinforcement learning from human feedback (RLHF)\cite{ouyang2022training} has been introduced to align model outputs with standards and ethical considerations. These approaches ensure that LLMs generate not only coherent but also scientifically accurate and ethically appropriate responses, making them increasingly suitable for applications across various domains including data analysis, literature review, hypothesis generation, and interdisciplinary research.

State-of-the-art development frameworks, such as Lang\-Chain\cite{langchain}, enhance the integration of these advanced capabilities, enabling LLM-agents to interact seamlessly with sophisticated network monitoring tools and management systems. We break down the operational mechanism of LLM-agents with four key phases: (1) parsing and interpreting input from the environment, (2) formulating action plans through the decomposition of complex tasks, (3) executing these actions via specialized tools or APIs, and (4) evaluating outcomes to inform subsequent decisions. This autonomous cycle facilitates interaction with diverse systems, making LLM-agents particularly valuable for tasks requiring continuous monitoring and adaptive responses in fields such as experimental design, data interpretation, and hypothesis generation.

\section{OFCnetLLM Design}
Figure \ref{arch} shows the OFCnetLLM design broken into the following principles:

Contemporary GPT-based LLMs demonstrate significant computational capabilities. However, processing complex computational tasks and analyzing high-dimensional data through a single model instance presents substantial methodological challenges and increases the probability of stochastic variations and hallucinations. Our methodology implements a distributed reasoning architecture utilizing multiple specialized agents that systematically identify network monitoring problems, decompose large-scale datasets into processed subsets, and address computational challenges through a robust, efficient, and fault-tolerant approach.

The automation of network data stream monitoring and corresponding response protocols encompasses routine computational operations that can be executed autonomously. We have implemented an optimized data processing pipeline incorporating LLM-based analytical methods for anomaly detection and monitoring optimization. The system's LLM-powered interface is also readily available to take queries, understand their intent, and perform appropriate actions.

The operation of complex LLM architectures necessitates substantial computational resources. Many implementations depend on proprietary LLM models that require network transactions, data processing, and storage off-premises, which can raise security concerns. To maintain optimal security protocols, local hosting and execution of the LLM model is essential. We selected Llama 3.2, a 2-billion parameter model known for its computational efficiency in the AI and machine learning community. This model enables operation on a single workstation with consumer-grade GPU acceleration, allowing OFCnetLLM to run quickly while maintaining a lightweight footprint.

\subsection{Reasoning}

Reasoning constitutes a systematic methodology for augmenting the capabilities of Large Language Model-based systems, enabling the resolution of complex real-world problems. The implementation of OFCnetLLM employs a multi-stage reasoning framework that replicates the cognitive processes demonstrated by human operators during network monitoring operations through the following steps.

\begin{enumerate}
    \item Monitoring: Systematically process network data str-eams into processed subsets that characterize network parameters and anomalies
    \item Identification: Execute classification to determine data segments requiring additional analytical evaluation
    \item Solution: Implement comprehensive data analysis protocols utilizing specialized computational tools
\end{enumerate}

The implementation of multi-stage reasoning chains is facilitated through LangChain, a specialized framework that helps the development of LLM-agent architectures and implementing of Chain-of-Thought reasoning methodologies, which is a fundamental component of cognitive processing protocols.

\subsection{Automation}

A fundamental characteristic of mission-critical systems is the systematic monitoring and quantitative assessment of real-time data streams for timely and accurate responses. The operational framework of OFCnetLLM implements continuous data stream monitoring, computational analysis of temporal data patterns, and predictive modeling of subsequent data batch events. The system architecture enables precise specification of analytical time windows to accommodate mission-critical application requirements.

While the monitoring functions operate autonomously, OFCnetLLM facilitates human-system interaction via a query interface. The language model component processes and interprets user-initiated queries, executing corresponding in-depth analysis of the network dataset. The model maintains a computational history of previous transactions and analytical procedures, thereby optimizing operational efficiency and minimizing cognitive load required for query formulation.

\section{Experiment Details}
\begin{figure*}[t!]
  \centering
  \begin{subfigure}[t]{0.48\textwidth}
    \centering
    \includegraphics[width=\linewidth,keepaspectratio]{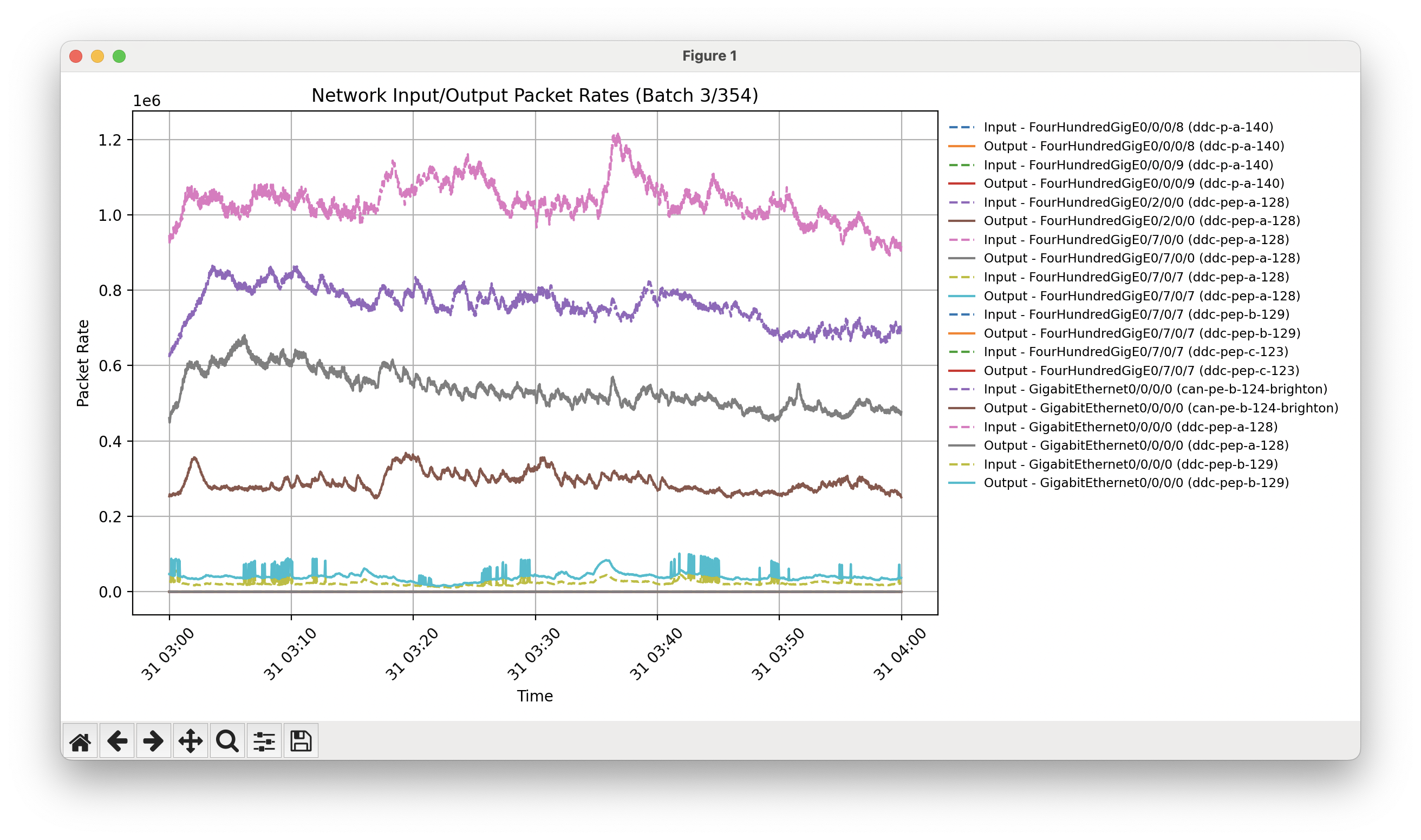}
    \caption{Batch 1.}
  \end{subfigure}\hfill
  \begin{subfigure}[t]{0.48\textwidth}
    \centering
    \includegraphics[width=\linewidth,keepaspectratio]{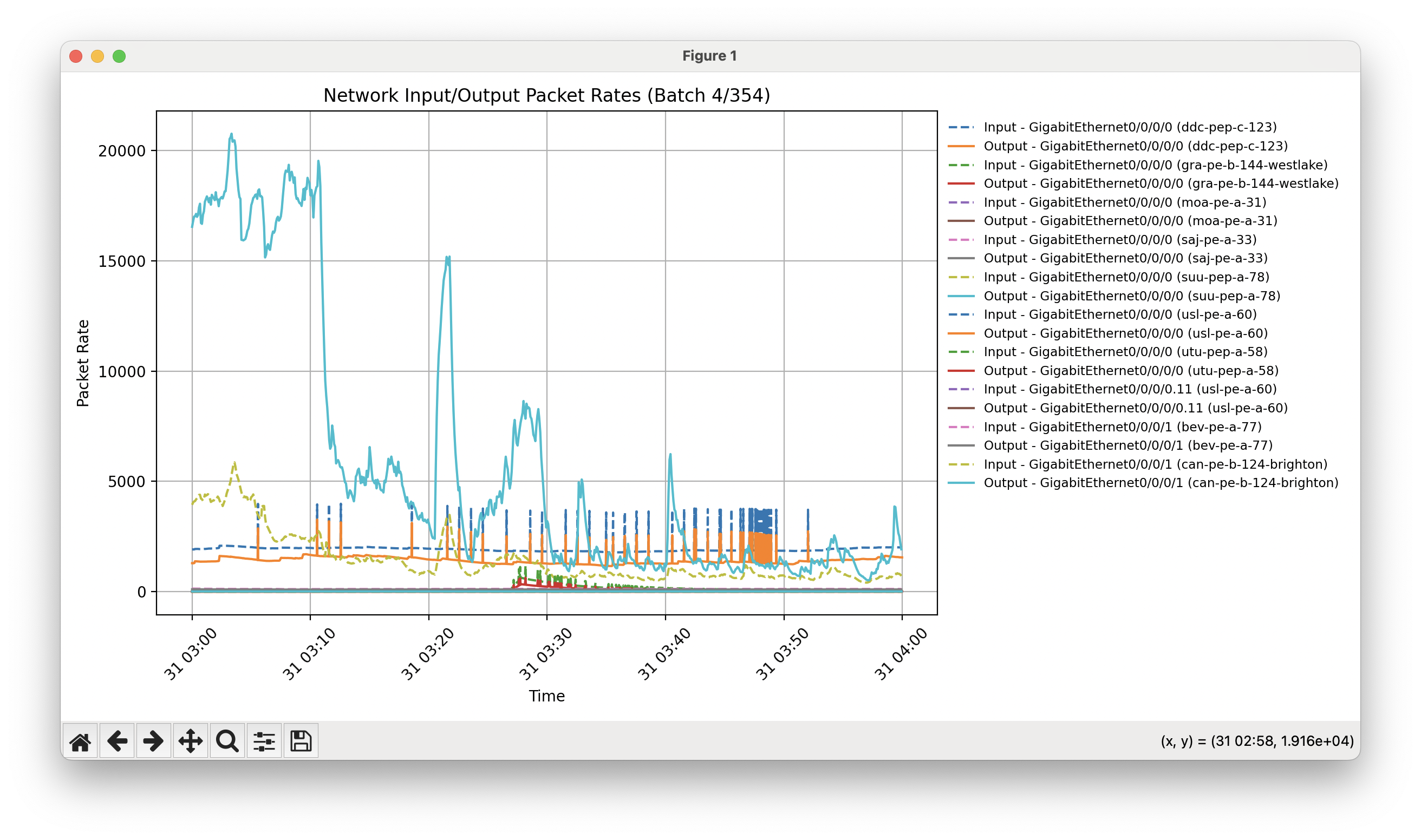}
    \caption{Batch 2.}
  \end{subfigure}
  \caption{Network interface data used for training OFCnetLLM}
  \label{train}
\end{figure*}

OFCnetLLM was demonstrated at the Optical Fiber Communication Conference (OFC2025) as part of the show floor network, performing live monitoring of data collection. Utilizing 2024 datasets, the model was trained on 19,108,243 unique data points gathered throughout the conference (approximately 13.4 million interface data points and 623K optical data points). These were consolidated into 1 million samples for training the ML model to predict network traffic through hourly timeframe analysis. The dataset encompassed network packet rates, input/output error rates, interface specifications, flow data, and live demonstrations from exhibition booths. Representative examples of these datasets are illustrated in Figure \ref{train}.

\begin{figure*}
\centering
\includegraphics[height=2in, width=5in]{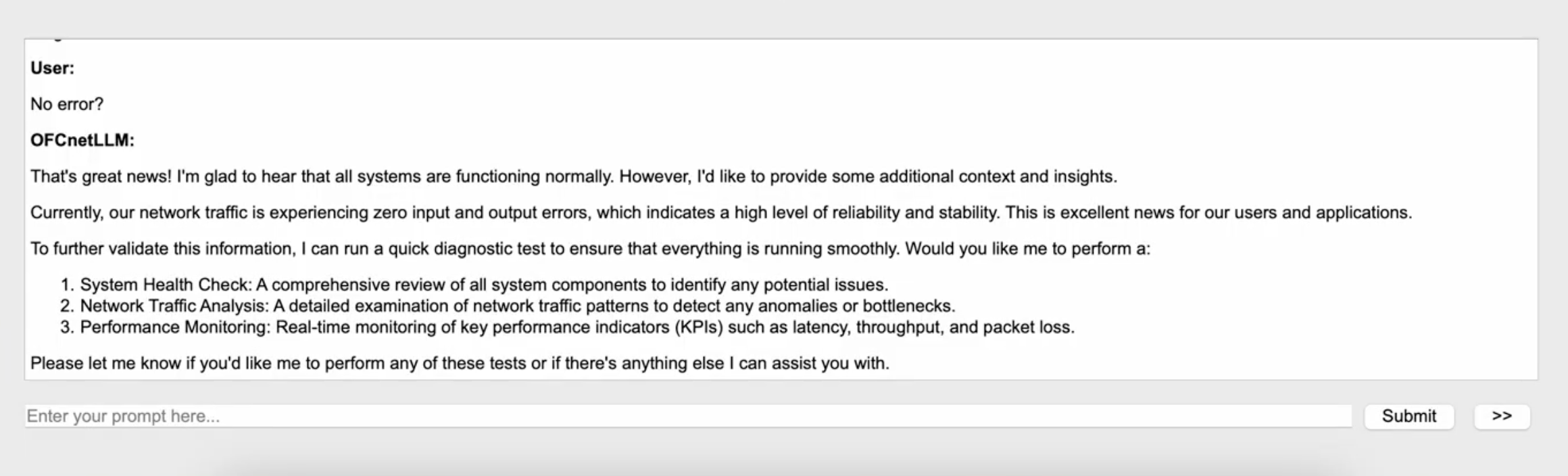}
\caption{Sample GUI for OFCnetLLM}\label{input}
\end{figure*}

\begin{figure*}
\centering
\includegraphics[height=3in, width=6in]{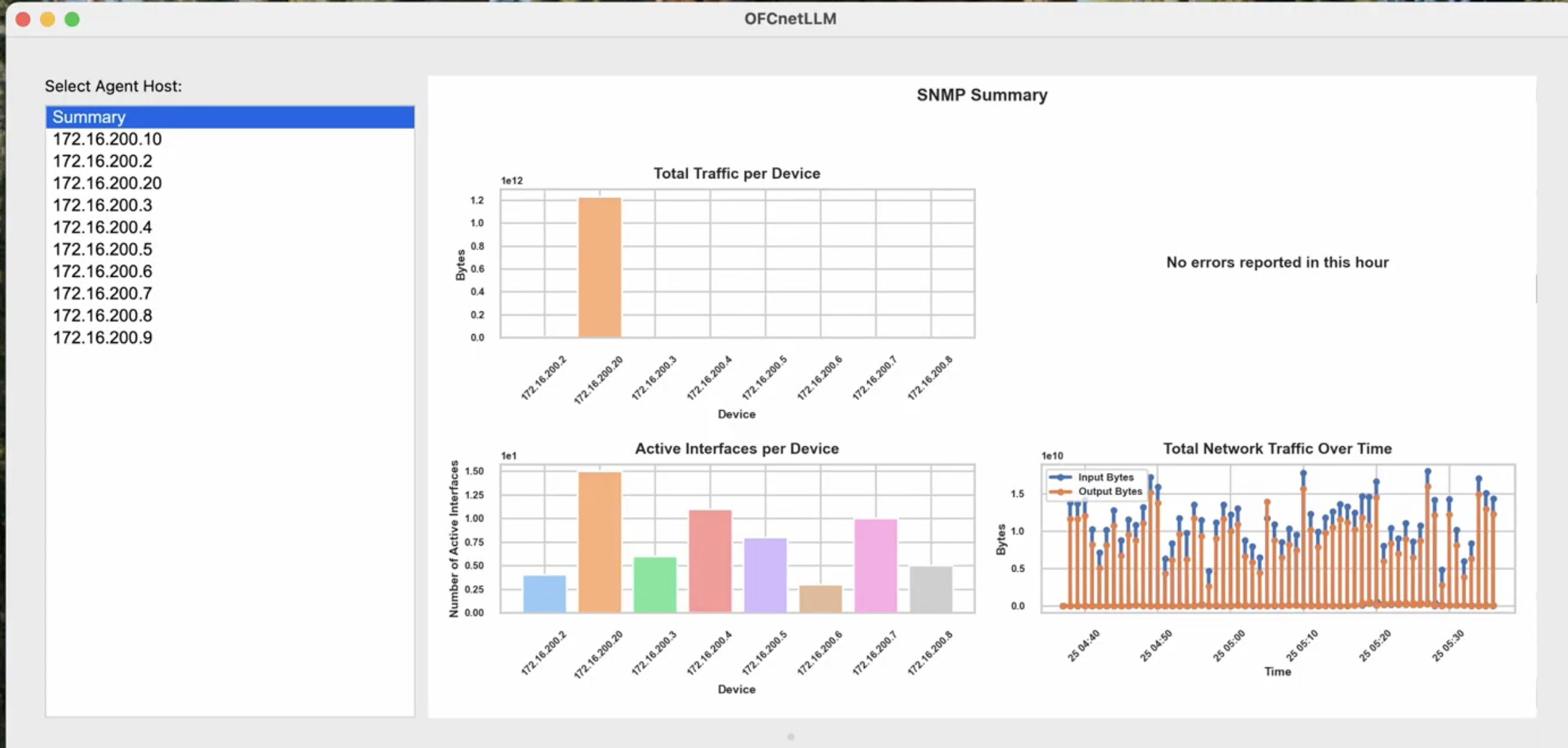}
\caption{Summary of all interfaces connected.}\label{sum}
\end{figure*}

\begin{figure*}
\centering
\includegraphics[height=3in, width=6in]{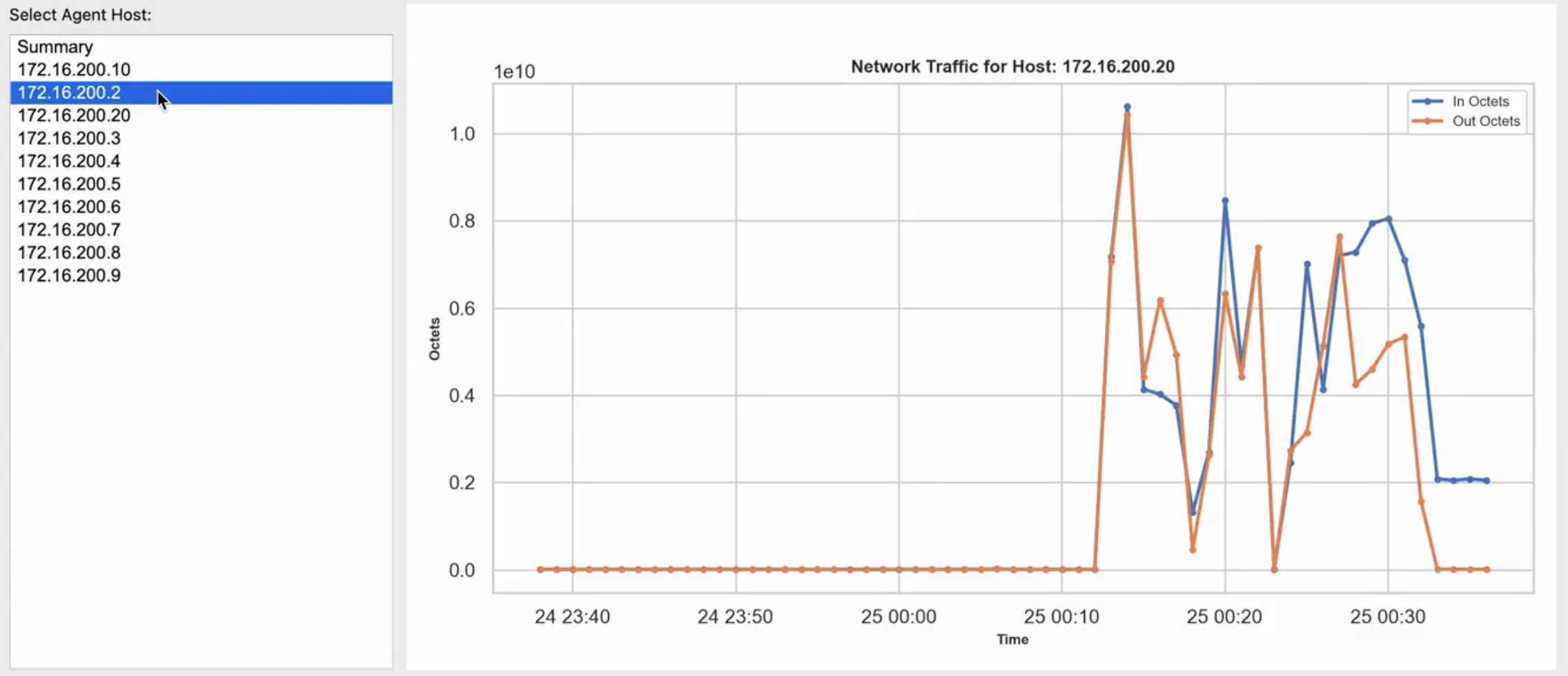}
\caption{Diagnosing individual interfaces.}\label{ind}
\end{figure*}

OFCnetLLM used a chat box to allow engineers to engage with the data set (Figure \ref{input}) producing outputs (Figure \ref{sum}, \ref{ind})

\section{Future Work and Conclusions}
OFCnetLLM was an LLM designed for OFC Network Monitoring. Initially this was developed over ChatGPT, as this is open-source code that is easy to adapt, but because of data security concerns raised by companies presenting at OFC wanting to keep their data private, this version was scrapped. As a result, a new LLM model based on a small Llama 3.2 was developed, trained on last year's OFC data. This model included one-shot training and reasoning to help motivate fault localization. Here, each agent was responsible for specific databases, and once a pattern was seen emerging in one database, the agent could talk to other agents and find patterns being seen in other databases. This can help AI agents use chain-of-thought to help diagnose problems, saving time for the engineers traversing through all databases to localize a fault.

The multiple agents managing multiple databases not only introduced efficiency but also helped develop security protocols for each database. More work is needed here, but each demonstrator was able to have their own agents, and only agents were interacting with each other while running on a local node, keeping the data secure and unshared. This allows for future research on how we can build efficient database management systems where local databases are not shared.

While OFCnetLLM was trained on last year's data, it will be enhanced with future OFC data. This allows the LLM to have more data available so it can learn to recognize patterns, allowing better prediction and diagnosis in the future.

Essentially, we found that this approach helps us design better multi-agent systems that can help improve queries for heterogeneous database design. There is further research needed on how these kinds of monitoring systems can help build better networks and how engineers will interact with these systems in their daily work.

\section*{Acknowledgments}

This manuscript has been authored by UT-Battelle, LLC, under contract DE-AC05-00OR22725 with the US Department of Energy (DOE). The US government retains and the publisher, by accepting the article for publication, acknowledges that the US government retains a nonexclusive, paid-up, irrevocable, worldwide license to publish or reproduce the published form of this manuscript or allow others to do so, for US government purposes. DOE will provide public access to these results of federally sponsored research in accordance with the DOE Public Access Plan (https://www.energy.gov/downloads/doe-public-access-plan). 

This work was supported by the U.S. DOE Office of Science, Office of Advanced Scientific Computing Research Early Career Grant ``Large Scale Deep Learning for Intelligent Networks'' award ERKJ435 hosted at Oak Ridge National Laboratory and partially by SWARM project grant number DE-SC0024387. 

This work was done in collaboration with the OFCnet Security and Measurements team at OFC2025 and we will like to thank Sana Bellamine and OFCnet members for allowing us to develop this work on their data sets.

\bibliographystyle{unsrt}  
\bibliography{references}

\end{document}